\documentstyle[e-e-ijmpa,twoside]{article}
\input epsf 
\def\DESepsf(#1 width #2){\epsfxsize=#2 \epsfbox{#1}}
\def\emem{e^-e^-}
\def\emg{e^-\gamma}
\def\eg{e\gamma}
\def\gev{\rm GeV}
\def\ot{\omega_T^{}}
\def\Mo{M_{\ot}}

\def\wlwl{W_L^-W_L^-}
\renewcommand{\thefootnote}{\fnsymbol{footnote}}

\begin{document}
\normalsize\textlineskip
\pagestyle{empty}

\font\fortssbx=cmssbx10 scaled \magstep1
\hbox to \hsize{
\includegraphics{uwlogo.ps}
\hskip.25in \raise.05in\hbox{\fortssbx University of Wisconsin - Madison}
\hfill$\vcenter{\normalsize\hbox{\bf MADPH-98-1049}
                \hbox{\bf hep-ph/9803358}
                \hbox{March 1998}}$ }

\title{ STRONGLY-INTERACTING ELECTROWEAK SECTOR\\
AT \boldmath$e^-e^-$ COLLIDERS\footnotemark}
\footnotetext{$^*$Talk presented
at the 2$^{nd}$ International Workshop on $\emem$ Interactions
at TeV Energies, UC-Santa Cruz, September 22-24, 1997.}

\author{ TAO HAN%
\footnote{ Work supported by the US Department of Energy under
Contracts DE-FG03-91ER40674 and DE-FG02-95ER40896.}
}

\address{Department of Physics, University of California,
Davis, CA 95616, USA\\
and\\
Department of Physics, University of Wisconsin,
Madison, WI 53706, USA}

\maketitle\abstracts{
We study the possible experimental signatures resulting 
from a strongly-interacting electroweak sector at $e^-e^-$
colliders, emphasizing the signal enhancement by high
beam polarization. We also discuss the unique role
for operating the collider in $e^-\gamma$ mode to produce
a heavy isosinglet vector state $\ot$.}

\setcounter{footnote}{0}
\renewcommand{\thefootnote}{\alph{footnote}}

\vspace*{1pt}\textlineskip
\section{Introduction}

It has been pointed out that a TeV  $\emem$ linear collider 
may be unique in probing new physics in the weak isospin
$I=2$ channel for the longitudinal $WW$ 
scattering\cite{emem,epem} $W_L^-W_L^- \to W_L^-W_L^-$.
Even if there is no resonance
in this channel, a strongly-interacting electroweak sector 
(SEWS) may still yield a sufficiently large signal rate. However, the 
continuum standard model (SM) backgrounds are substantial
and sophisticated kinematical cuts are needed to isolate
the signal. On the other hand, as anticipated, an $\emem$ linear
collider may achieve high polarization for the $e^-$ beams.\cite{pol} 
This will lead to significant improvement for the signal 
since not only the signal rate will be enhanced
but also some background processes can be suppressed by 
properly choosing the beam polarization. We study this
question quantitatively in Sec.~2.

Another attraction for an $\emem$ collider is the possibility
of operating it in $\emg$ mode by the back-scattering
of a low-energy laser beam.\cite{telnov} 
This may provide a unique opportunity
to produce a heavy isosinglet vector state $\ot$
in a strongly-interacting electroweak sector. 
We study the $\ot$ signal and corresponding
backgrounds in Sec.~3. We summarize our
discussions in Sec.~4.


\section{SEWS Signal at \boldmath$\emem$ Colliders with High Beam Polarization}

If no light Higgs boson is found for $m_H^{}$ to be less than 
about 800 GeV,  one would
anticipate that the interactions among longitudinal vector bosons
become strong.\cite{chan-gail} 
Without knowing the underlying dynamics for the
strongly-interacting electroweak sector (SEWS), 
we will have to parametrize the physics by an 
effective theory, with possible low-lying resonant
states.\cite{baggeretal}

The simplest model for a strongly-interacting 
$W_L^-W_L^-$ sector is the exchange of a heavy scalar 
(Higgs) boson. This results in an enhancement of the
$e^-e^-\to\nu\nu W^-W^-$ production cross section compared to that
expected from the exchange of a light Higgs boson in the SM. 
This enhancement
due to a Higgs boson of mass 1~TeV can be defined as the difference of
the $W_L^-W_L^-\to W_L^-W_L^-$ fusion contributions
\begin{equation}
\Delta\sigma_H = \sigma(m_H=1~{\rm TeV}) - \sigma(m_H=0.1~\rm TeV)
\end{equation}
to $e^-e^-\to\nu\nu W^-W^-$ production. There is no appreciable
numerical change between the choices $m_H=0.1$~TeV and $m_H=0$ for the
light Higgs boson reference mass. We find the values\cite{emem,epem}
\begin{equation}
 \Delta \sigma_H \simeq \left\{
\begin{array}{ll}
53.6 - 50.9 = 2.7~{\rm fb}
& \mbox{at $\sqrt s = 1.5$~{\rm TeV}; } \\
86.5 - 82.0 = 4.5 ~{\rm fb}
& \mbox{at $\sqrt s = 2$~{\rm TeV}. }
\end{array}
\right.
\label{eq:rate}
\end{equation}
This implies that the signal rate at a high luminosity 
collider with 100-200 fb$^{-1}$/yr is quite sizeable. 
However, the total cross section 
for $e^-e^-\to\nu\nu W^-W^-$ from all Standard-Model
diagrams (dominantly from the transversely polarized
gauge bosons $W_T^- W_T^-$) is about 20 times larger than
$\Delta\sigma_H$. Hence the background contributions 
associated with $W_T^-W_T^-$, $W_T^-W_L^-$ must somehow be
selectively reduced if we are to observe the
strongly-interacting $W_L^-W_L^-$ signal. Moreover,
when considering the $W$ hadronic decay, 
processes such as $e^-e^-\to e^-e^-W^+W^-$
and $e^-e^-\to e^-\nu W^-Z$ would also become potential
backgrounds, and one will have to scrutinize the backgrounds
carefully.
There are several ways to accomplish the substantial
background suppression. We first present the implementation
of selective kinematical cuts to isolate the signal. We
then study the improvement by employing beam polarization.

There are characteristic kinematical features for
the signal events in $W_L^-W_L^-$ 
scattering.\cite{wpwp,baggeretal,emem} For instance:
\begin{itemize}
\item
The signal gives large $M(W^-W^-)$ of order 1~TeV
with centrally-produced $W^-$ having large $p_T(W)$.
The two $W_L^-$'s in the final state are largely
back-to-back, resulting in large transverse
momentum difference 
$\Delta p_T(WW) = |{\bf p}_T(W_1)-{\bf p}_T(W_2)|$.
\item
The $p_T(WW)$ spectrum of the signal is peaked
around $M_W$ and falls off rapidly at high $p_T$ like $1/p_T^4$.
\item
Rejecting energetic electrons in the final state
will help remove the potentially large background
processes such as $e^-e^-\to e^-e^-W^+W^-$
and $e^-e^-\to e^-\nu W^-Z$, which are especially
dangerous when identifying $W_L^-$'s via the hadronic
mode.
\item
When including the finite jet-energy resolution
in considering the $W$ hadronic decay $W \to jj$, 
we adopt the energy smearing\cite{jlc}
$\delta E_j/E_j = 0.50 \Big/ \sqrt{E_j} \;\oplus\; 0.02 \; .$
It turns out to be feasible to discriminate the $W$ from $Z$ 
in the di-jet mode by their mass difference.
\end{itemize}
\begin{table}[tb]
\centering
\tcaption{Kinematical cuts and hadronic $W$ identification.}
\label{cuts}
\medskip
\setlength{\tabcolsep}{5pt}
\begin{tabular}{|l|c|} \hline
kinematical variable & selective cut \\ 
\hline\hline
$M_{WW}^{min}$ & 500 $\gev$\\
$p_T^{min}(W)$ & 150 $\gev$\\
$|\cos\theta_W^{max}|$ & 0.8 \\
$\Delta p_T^{min}(WW)$ & 400 $\gev$\\
$p_T^{}(WW)$  & 50 -- 300 $\gev$\\ \hline
electron veto & $E_e>50~\gev$  \\ 
in the range  & $|\cos\theta_e|<|\cos(150~{\rm mrad})|$ \\ \hline
$M(W\to jj)$  & $\left[0.85M_W, \; {1\over 2}(M_W+M_Z)\right]$  \\
$M(Z\to jj)$  & $\left[{1\over 2}(M_W+M_Z), \; 1.15M_Z\right]$  \\
\hline
\end{tabular}
\end{table}
We summarize our acceptance cuts in Table~\ref{cuts}. Before
including the hadronic decay branching fraction and 
the $M(W\to jj)$ reconstruction, 
the signal rate with the cuts becomes about 1.0 fb
at $\sqrt s = 2$ TeV, where the remaining backgrounds 
are  2.8~fb for $e^-e^-\to \nu\nu W_T^-W_T^-$, 
4.4~fb for $e^-e^-\to e^-e^- W^-W^+$ 
and 4.7~fb for $e^-e^-\to e^-\nu W^-Z$.
At $\sqrt s=1.5$ TeV, the signal rate is down to
about 0.4 fb,
making the signal observation more difficult. 
Including the $W/Z$ discrimination through
the di-jet mass of their decay products improve the
signal-to-background ratio significantly. 
Monte Carlo simulation\cite{epem}
indicates that true $W W$, $W Z$, $ZZ\to jjjj$ 
events will be interpreted statistically as follows:
$$
\begin{array}{lcrrrrr}
WW &\Rightarrow 
& 73\%\: WW, & 17\%\: WZ, &  1\%\: ZZ,&  9\%\: {\rm reject},
\\
WZ &\Rightarrow 
& 19\%\: WW, & 66\%\: WZ, & 7\%\: ZZ,&  8\%\: {\rm reject},
\\
ZZ &\Rightarrow 
&  5\%\: WW, & 32\%\: WZ, & 55\%\: ZZ,&  8\%\: {\rm reject}.
\end{array}
$$
By beating down the persistent $WZ$ background which
escapes the ``electron veto'' cut,
this helps improve the signal observability.

Applying the cuts and signal efficiency obtained from the
above study based on the heavy scalar model, 
we can estimate the signal rates for other SEWS
scenarios.\cite{baggeretal,epem}
We consider a chirally-coupled scalar boson ($m_S=1$ TeV
and $\Gamma_S=350$ GeV), a chirally-coupled vector boson ($m_V=1$ TeV
and $\Gamma_V=25$ GeV), and the low energy
theorem amplitude. These calculations are
carried out with the effective $W$-boson approximation,
which is justified for the current 
application.\cite{wpwp,baggeretal}

\begin{table}[tb]
\centering
\tcaption{
$W^-_LW^-_L$ signals for different
models of strongly-interacting $W$ sector
at an $e^-e^-$ collider for $\sqrt s=1.5$~TeV, 
with cuts listed in Table~\ref{cuts}. Polarization effects
for $P_e$=0, $-85\%$ and $-100\%$ (both electron beams ) are compared.
Backgrounds are summed over $W^-W^-$ with a light Higgs
exchange, $W^+W^-$ and $W^-Z$. 
Entries correspond to the number of events with hadronic $W,Z$
decays for an integrated luminosity of 200 fb$^{-1}$. 
$W/Z$ identification via di-jet mass has been implemented,
as discussed in
the text to improve the signal/background ratio.
As a rough indication of the signal observability, 
values of $S/\protect \sqrt B$ are also given.}
\label{table2}
\medskip
\begin{tabular}{|l|c|c|c|c|c|c|} \hline
 $\protect \sqrt s=1.5$ TeV & SM  & Scalar & Vector   & LET & Bckgnds \\
$M_{WW}^{min}$=0.5 TeV & $m_H=1$ TeV & $m_S=1$ TeV & $m_V=1$ TeV & & \\
\hline\hline
 $P_e=0$ & 27  & 35   & 36 & 42  & 230  \\
$S/\protect \sqrt B$
& 1.8  & 2.3  & 2.4 & 2.8 & {} \\ \hline
 $P_e=-85\%$ & 93  & 121  & 123 & 144 & 620  \\
$S/\protect \sqrt B$
& 3.8  & 4.8  & 5.0 & 5.8 & {} \\ \hline
 $P_e=-100\%$ & 109  & 141   & 144 & 168  & 713  \\
$S/\protect \sqrt B$
& 4.1  & 5.3  & 5.4 & 6.3 & {} \\ \hline
\end{tabular}
\end{table}

The predicted numbers of events with hadronic $W$ decay 
at $\sqrt s = 1.5$~TeV are presented in Table~\ref{table2}. 
Results are given with the cuts listed in Table~\ref{cuts} 
and for an integrated luminosity of 200~fb$^{-1}$.  
So far, we have ignored the $e^-$-beam polarization
and the corresponding results are in the first row
of Table~\ref{table2} ($P_e=0$). We see that the 
statistical significance is rather poor for all models, 
barely reaching a 3$\sigma$ effect 
for the best signal of the LET model. 

\begin{table}[tbh]
\centering
\tcaption{
Same as Table~\ref{table2}, but with
an integrated luminosity of 300 fb$^{-1}$.}
\label{table3}
\medskip
\begin{tabular}{|l|c|c|c|c|c|c|} \hline
 $\protect \sqrt s=1.5$ TeV & SM  & Scalar & Vector   & LET & Bckgnds \\
$M_{WW}^{min}$=0.5 TeV & $m_H=1$ TeV & $m_S=1$ TeV & $m_V=1$ TeV & & \\
\hline\hline
 $P_e=0$ & 41  & 53   & 54 & 63  & 345  \\
$S/\protect \sqrt B$
& 2.2  & 2.8  & 2.9 & 3.4 & {} \\ \hline
 $P_e=-85\%$ & 140  & 181  & 185 & 216 & 930  \\
$S/\protect \sqrt B$
& 4.6  & 5.9  & 6.1 & 7.1 & {} \\ \hline
 $P_e=-100\%$ & 164  & 212   & 216 & 252  & 1070  \\
$S/\protect \sqrt B$
& 5.0  & 6.5  & 6.6 & 7.7 & {} \\ \hline
\end{tabular}
\end{table}

However, high $e^-$-beam polarizations seem achievable 
at the NLC\cite{jlc,pol}
and one would like to ask its implication on SEWS physics.
If we denote the percentage of the longitudinal beam
polarization along the beam direction
by $P_e$, with $P_e=-1(+1)$ for $e^-_L(e^-_R)$,
we can express the scattering
matrix element squared for a given physical process by
\begin{eqnarray}
\overline\Sigma|{\cal M}_{tot}|^2
&=&{1\over 4}\Sigma_{\lambda_1,\lambda_2}|
{\cal M}(\lambda_1,\lambda_2)|^2\nonumber \\
&=&{1\over 4}[\  (1+P_{e1})(1+P_{e2})|{\cal M}(+,+)|^2\nonumber \\
\ &&+\ (1+P_{e1})(1-P_{e2})|{\cal M}(+,-)|^2\\ 
\ &&+\ (1-P_{e1})(1+P_{e2})|{\cal M}(-,+)|^2\nonumber \\
\ &&+\ (1-P_{e1})(1-P_{e2})|{\cal M}(-,-)|^2\nonumber \ ],
\end{eqnarray}
where ${\cal M}(\lambda_1,\lambda_2)$ is the
helicity amplitude with initial state electronic helicities
$\lambda_1$ and $\lambda_2$. Since the $\wlwl$
scattering signal is through purely left-handed 
currents in ${\cal M}(-,-)$,
while some dominant backgrounds such as $\gamma\gamma \to W^+W^-$
and $WZ$ final state are non-chiral, one could significantly
improve the signal observability by employing a left-handed
longitudinally polarized $e^-$ beam. 
As an illustration, we consider $P_e=P_{e1}=P_{e2}=-85\%$. 
Then the matrix element squared becomes
\begin{equation}
\overline\Sigma|{\cal M}_{tot}|^2 \approx
0.0056\ |{\cal M}(+,+)|^2 +
0.14\ |{\cal M}(+,-)|^2 + 0.86\ |{\cal M}(-,-)|^2.
\end{equation}
The numerical coefficients in front of the helicity amplitudes
squared
clearly demonstrate the advantage for choosing the left-handed
beam polarizations. In the ideal case with $P_e=-100\%$,
the signal cross section from ${\cal M}(-,-)$ would be
enhanced by a factor of 4 with respect to the unpolarized one, 
while the backgrounds involving
$e^-_R$ would be eliminated. Our corresponding results
are shown in Table~\ref{table2}, indicated by $P_e=-85\%$ and
$P_e=-100\%$. We see that the signal statistical significance
is essentially doubled by employing $-85\%$ beam polarizations.
In Table~\ref{table3}, results for 300 fb$^{-1}$ are shown.

\section{\boldmath$\omega_T^{}$ Signal in \boldmath$\eg$ Collisions}

\begin{figure}[b]
\centering
\leavevmode
\epsfxsize=4.8in\epsffile{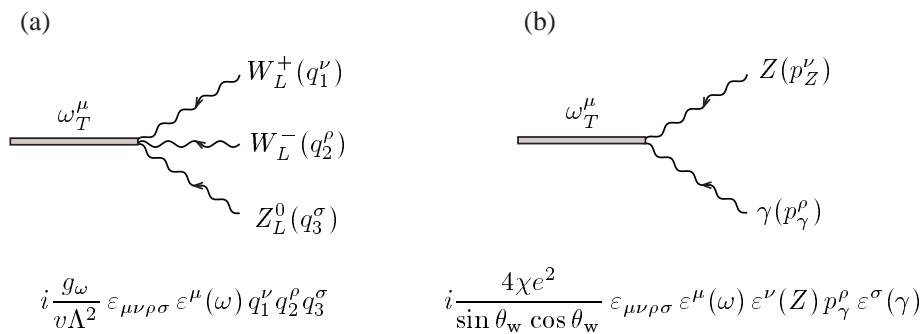}

\caption[]
{Effective interactions of $\ot$ with (a)
$W_L^+W_L^-Z_L$ and (b) $Z\gamma$.}
\label{vertex}
\end{figure}

In many dynamical electroweak symmetry breaking models,
it is quite common that there exist other resonant
states\cite{tc} besides 
an isotriplet vector ($\rho^{}_T$) 
and an isosinglet scalar ($H$), 
such as an isosinglet vector $\ot$ 
and isotriplet axial vector $A^{}_T$. 
In fact, it has been argued
that to preserve good high energy behavior in a SEWS
sector, it is necessary for {\it all} the above 
resonant states to coexist.\cite{hhh}
It is therefore wise to keep
an open mind and to include other characteristic
resonant states in examining the SEWS physics at colliders.

We concentrate on the isosinglet vector $\ot$. It couples
to three longitudinal gauge bosons 
(electroweak Goldstone bosons)
strongly and to $Z\gamma$ electroweakly. 
The interactions can be parametrized
effectively by two parameters $\chi$ and 
$g_{\omega}/\Lambda^2$, with $\chi,g_{\omega}$ naturally
order of one and $\Lambda$ the new physics scale
in the SEWS sector, typically $\Lambda \le 4\pi v\approx 3$
TeV. More explicitly, we assume the couplings as
those in Fig.~\ref{vertex}. It is easy to see that
one can trade the coupling parameters to the two
partial widths $\Gamma(WWZ)$ and $\Gamma(Z\gamma)$.
Including the mass $\Mo$,
there are three physical parameters in this sector.

\begin{figure}[tb]
\centering
\leavevmode
\epsfxsize=4.0in\epsffile{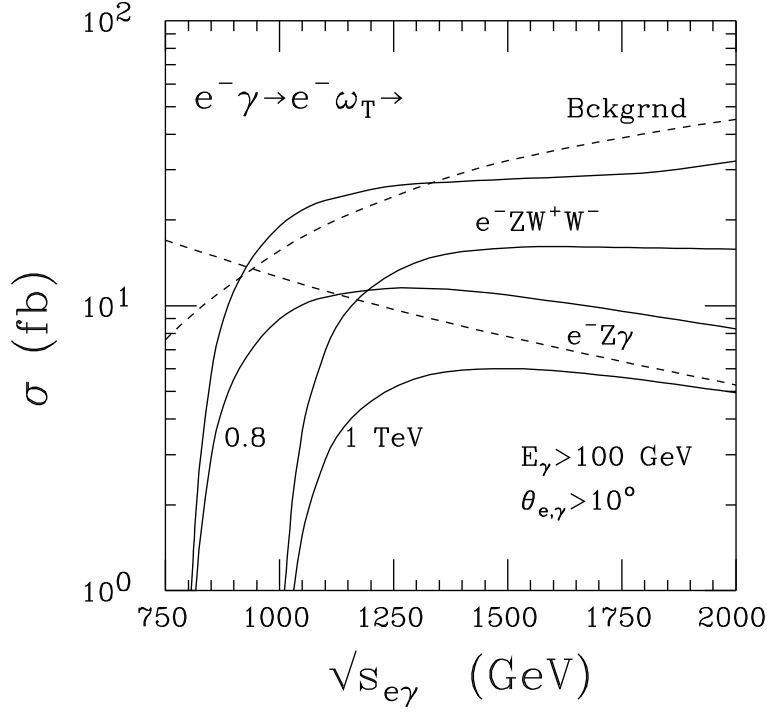}

\caption[]{
Cross sections versus the $\eg$ c. m. energy
for the signal $\emg \to e^-\ot$ with
$\ot \to W^+W^-Z$ and $Z\gamma$ (solid curves), 
and the SM backgrounds (dashes). 
Results for two mass values, $M_{\ot}=0.8$ and
1.0 TeV, are presented.}
\label{sigom}
\end{figure}

\begin{figure}[tb]
\centering
\leavevmode
\epsfxsize=4.5in\epsffile{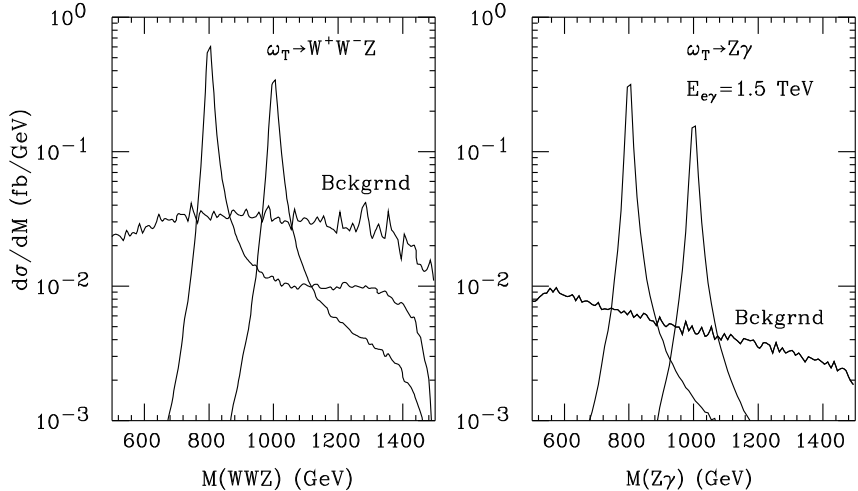}

\caption[]{
Differential cross sections as a function of
the invariant mass of the $\ot$ decay products
$M(WWZ)$ and $M(Z\gamma)$.
for the signal $\emg \to e^-\ot$ with
$\ot \to W^+W^-Z$ and $Z\gamma$ (solid curves), 
and the SM backgrounds (dashes). 
Results for two mass values, $M_{\ot}=0.8$ and
1.0 TeV, are presented.}
\label{mom}
\end{figure}

The direct $\ot Z\gamma$ coupling in Fig.~\ref{vertex}(b)
implies that an $\ot$ can be effectively produced by
$Z\gamma$ fusion in $\eg$ collisions. This is indeed
quite a unique feature for an $\emem$ linear collider
operating in $\eg$ mode. As an exploratory study, 
we choose the parameters as follows
$$
\begin{array}{ccc}
\Mo\ [{\rm TeV}] & \Gamma(WWZ)\  [\gev] & \Gamma(Z\gamma)\  [\gev] \\
 0.8            & 16                 & 4  \\
 1.0            & 20                 & 5. \\
\end{array}
$$
In Fig.~\ref{sigom} we show the signal cross 
section versus the c.m. energy of an $\eg$ collider
for both modes $\ot \to W^+_LW^-_LZ_L$ and $Z\gamma$
by the solid curves. The corresponding SM backgrounds
are also presented by the dashed curves. For the $WWZ$
channel, the energy
dependence of the $\ot$ coupling to $ W^+_LW^-_LZ_L$ 
gives the rise for the signal cross section at higher
energies; while the contribution from $\gamma^* \to W^*W$
makes the SM background cross section for $\emg \to e^-ZW^+W^-$
increase with $\sqrt{s^{}_{\eg}}$.
The cross sections for $\emg \to e^-Z\gamma$ 
process (both signal and background)
at a finite angle $\theta_e,\theta_\gamma>10^\circ$ decrease
at higher $\sqrt{s^{}_{\eg}}$. It appears from the figure
that the SM backgrounds are always at least as large
as the signals in the total rate. However, we should notice that the
signal is a resonant production and invariant mass
spectra for the final states should reconstruct $\Mo$.
This is demonstrated in  Fig.~\ref{mom} where the
signal resonant peaks near  $\Mo=0.8,\ 1.0$ TeV
are clearly observable over the continuum SM backgrounds,
where $\sqrt{s^{}_{\eg}}=1.5$ TeV is assumed. The signal
rate near the peak is substantial, being more than
a few hundred events per 100 fb$^{-1}$.
Further studies of related physics are in progress.\cite{ghk}

\section{Summary}

We discussed the possibility of observing a strong $W^-_LW^-_L$ 
scattering signal from a strongly-interacting electroweak 
sector (SEWS), 
which occurs through the weak isospin $I=2$ channel and is unique for
$e^-e^-$ collisions. We summarized the necessary kinematical cuts to
significantly reduce the $W^-_TW^-_T$, $W_T^-W_L^-$, $W^+W^-$ and $W^-Z$ 
backgrounds to the $W_L^-W_L^-$ signal in hadronic $W$ decay mode.
We quantified the effects of the beam polarizations and demonstrated
the significant improvement for the signal observability
by employing left-handed beam polarizations.
We also pointed out the unique role for a TeV $\eg$ collider
in producing other resonant states in SEWS, such as
an isosinglet vector state $\ot$. 

The interesting physics opportunity at a TeV $\emem$ collider 
as well as $\eg$ and $\gamma\gamma$ colliders should be further explored.

\vskip1.5pc
\leftline{\bf Acknowledgments}

\vskip6pt

I would like to thank Clem Heusch and colleagues
at UC-Santa Cruz for organizing this 
stimulating workshop.

\newpage


\end{document}